\documentstyle[12pt]{article}

\author{Hao Chen\\
Department of Mathematics\\
Zhongshan University\\
Guangzhou,Guangdong 510275\\
People's Republic of China\\
and\\
Department of Computer Science\\
National University of Singapore\\
Singapore 117543\\
Republic of Singapore}
\title{New Invariants and Separability Criterion of the Mixed States: Bipartite Case}
\date{July,2001}

\begin{document}

\maketitle
\begin{abstract}
We introduce algebraic sets in the complex projective spaces for the mixed states in bipartite quantum systems as their invariants under local unitary operations. The algebraic sets have to be the union of the  linear subspaces if the mixed state is separable, and thus we give a new criterion of separability. Some examples of the entangled mixed states are constructed and studied based on our criterion. Our invariants also can be used to distinguish inequivalent mixed states under local unitary operations.

\end{abstract}

In recent years it became clear that entanglement is one of the most important ingredients and resources of quantum infromation processing(see [1],[2]), and thus stimulated tremendous studies of quantum entanglements of both pure and mixed states from both theoretical and experimental view, for a survey we refer to [3],[8] and [9]. For mixed states, the criterion of Peres-Horodecki said that a separable mixed state must necessarily have a positive partial transpose(PPT) , and this is also a sufficent condition for separability of mixed states in $2 \times 2$ or $2 \times 3$ quantum systems( [5],[6]). It is also noted that the mixed entangled states with PPT cannot be distilled (bound entanglement). The first several examples of the entangled mixed states  with PPT are provided in [7]. The context of unextendible product bases was introduced in [10] and [11] as a systematic way to construct entangled mixed PPT states in both bipartite and multipartite case. For ``low''  rank PPT mixed states it is proved that they have to be separable for both bipartite and multipartite quantum systems ([12],[13],[14]). We also should mention the elegant geometric approach for the entanglement of pure states in n qubits case in [16], which leads further the concepts and results of Schmidt number and coefficients [17]. \\

In this letter, we introduce algebraic sets (ie., the zero locus of several multi-variable homogeneous polynomials, see [18]) in the complex projective space (see [18]) for any given mixed state in a bipartite quantum system with the following two properties:\\

1) When we apply local unitary operations to the mixed state, the corresponding  algebraic sets are changed by a linear transformation, and thus these invariants can be used to distinguish inequivalent mixed states under local unitary operations;\\

2)The algebraic sets are  linear (the union of some linear subspaces) if the mixed state is separable, and thus we give a new separability criterion.\\

From our Theorem 1 below, if the Fubini-Study metric of the complex projective space is used, the metric properties of these algebraic sets are also preserved when local  unitary operations are applied to the mixed state. Hence we establish a connection between Quantum Entanglement and both the Algebraic Geometry and Hermitian Geometry of these algebraic sets. Any algebraic-geometric or Hermitian geometric invariant of the algebraic set of the mixed state is an invariant of the mixed state under local unitary operations.(see [25]).\\

We only treat bipartite case in this letter and the multi-partite case will be treated in [23].\\

The algebraic sets used in this letter is called ``determinantal varieties'' in algebraic geometry ([18] Lecture 9 and [19] Cha.II) and have been studied by mathematicians from different motivations such as  geometry of curves ([19]), Hodge theory ([20]), commutative algebra ([21]) and even combinatorics ([22]). It is interesting to see that it can be useful even in quantum information theory. For the algebraic geomerty used in this letter and [23] we refer to the nice book [18].\\

Let $H=C_{A}^m \otimes C_{B}^n$ and the standard orthogonal base is $\{|ij>\}$, where, $i=1,...,m$ and $j=1,...,n$, and $\rho$ is a mixed state on $H$. We represent the matrix of $\rho$ in the base $\{|11>,...|1n>,...,|m1>,...,|mn>\}$, and consider $\rho$ as a blocked matrix $\rho=(\rho_{ij})_{1 \leq i \leq m, 1 \leq j \leq m}$ with each block $\rho_{ij}$ a $n \times n$ matrix corresponding to the $|i1>,...,|in>$ rows and the $|j1>,...,|jn>$ columns.\\

{\bf Definition 1.}{\em  We define 

$$
\begin{array}{ccccc}
V_{A}(\rho)=\{(r_1,...,r_m)\in CP^{m-1}:det( \Sigma_{i,j}r_ir_j^{*} \rho_{ij})=0\}
\end{array}
$$
Similarly $V_{B}(\rho) \subseteq C^n$ can be defined. Here * means the conjugate of complex numbers and det means the determinant.}\\

{\bf Example 1.} Let $\rho=\frac{1}{mn}I_{mn}$, the maximally mixed state, we easily have $V_{A}(\rho)=\{(r_1,...,r_n): det(\Sigma r_i r_i^{*}I_n)=0\}=\emptyset$. This is a trivial example.\\

{\bf Theorem 1.}{\em  Let $T=U_{A} \otimes U_{B}$, where $U_{A}$ and $U_{B}$ are the local operations (ie., unitary linear transformation) on $C_{A}^m$ and $C_{B}^n$ rescpectively. Then $V_{A}(T(\rho))=U_{A}^{-1}(V_{A}(\rho))$, that is $V_{A}(\rho)$ (resp. $V_{B}(\rho)$) is a ``invariant'' upto a linear transformation of $CP^{m-1}$ (resp. $CP^{n-1}$) of the mixed state $\rho$ under local unitary operations.}\\

{\bf Proof.} Let $U_{A}=(u_{ij}^{A})_{1 \leq i \leq m,1 \leq j \leq m}$, and $U_{B}=(u_{ij}^{B})_{1 \leq i \leq n, 1 \leq j \leq n}$, be the matrix in the standard orthogonal bases. Then the matrix of $T(\rho)$ under the standard orthogonal base $\{|ij>\}$, where $ 1 \leq i \leq m, 1 \leq j \leq n$, is $T(\rho)=(\Sigma_{l,k} u_{il}^{A} U_{B}(\rho_{lk})(U_{B}^{*})^{ \tau}(u_{jk}^{A})^{*})_{1\leq i \leq m, 1 \leq j \leq m}$. Hence \\

$$
\begin{array}{cccccccc}
V_{A}(T(\rho))=\{(r_1,...,r_m): det(\Sigma _{l,k} (\Sigma_i r_i u_{il}^{A})(\Sigma_i r_i u_{il}^{A})^{*}U_{B}(\rho_{lk})(U_{B}^{*})^{\tau})=0\}

\end{array}
(1)
$$

We set $r_l'=\Sigma_i r_i u_{il}^{A}$ for $l=1,...,m$.Then\\

$$
\begin{array}{ccccc} 
\Sigma _{l,k} (\Sigma_i r_i u_{il}^{A})(\Sigma_i r_i u_{il}^{A})^{*}U_{B}(\rho_{lk})(U_{B}^{*})^{\tau}\\
=U_{B}(\Sigma_{lk} r_l' (r_k')^{*} (\rho_{lk})(U_{B}^{*} )^{\tau}

\end{array}
(2)
$$

Thus\\
$$
\begin{array}{ccccccc}
V_A(T(\rho))=\{(r_1,...,r_m):det(\Sigma_{lk} r_l' (r_k')^{*} (\rho_{lk}))=0\}

\end{array}
(3)
$$

and our conclusion follows. \\

{\bf Remark 1.} Since $U_{A}^{-1}$ certainly preserve the Fubini-Study metric of $CP^{m-1}$, we know that all metric properties of $V_{A}(\rho)$ are preserved when the local unitary operations are applied to the mixed state $\rho$.\\

In the following statement, the term ``algebraic set `` means the zero locus of several multi-variable polynomials.(see [18]).\\

{\bf Theorem 2.} {\em $V_{A}(\rho)$ (resp. $V_{B}(\rho)$) is an algebraic set in $CP^{m-1}$ (resp. $CP^{n-1}$).}\\

{\bf Theorem 3.} {\em If $\rho$ is a separable mixed state, $V_{A}(\rho)$ (resp. $V_{B}(\rho)$) is a linear subset of $CP^{m-1}$ (resp. $CP^{n-1}$),ie., it is the union of the linear subspaces.}\\

For the purpose to prove Theorem 2 and 3 we need the following lemmas.\\

{\bf Lemma 1.} {\em We take the orthogonal base $\{e_1,...,e_h\}$ of a $h$ dimension Hilbert space $H$. Let $\rho=\Sigma_{l=1}^{t} p_l P_{v_l}$, where $v_l$ is unit vector in $H$ for $l=1,...,t$ ,   $v_l=\Sigma_{k=1}^{h} a_{kl} e_k$ , $A=(a_{kl})_{1\leq k \leq h, 1 \leq l \leq t}$ is the $h \times t$ matrix and $P_{v_l}$ is the projection to the vector $v_l$. Then the matrix of $\rho$ with the base $\{e_1,...,e_h\}$ is $AP(A^{*})^{\tau}$, where $P$ is the diagonal matrix with diagonal entries $p_1,...,p_h$.}\\

{\bf Proof}. We note that the matrix of $P_{v_l}$ with the base is $\alpha (\alpha^{*})^{\tau}$ where $\alpha=(a_{1l},...,a_{hl})^{\tau}$ is just the representation vector of $v_l$ with the base. The conclusion follows immediately.\\

The following conclusion is a direct matrix computation from Lemma 1 or see [5],[7].\\

{\bf Corollary 1.} {\em Suppose $p_i>0$, then the image of $\rho$ is the linear span of vectors $v_1,...,v_t$.}\\

Now let $H$ be the $C_{A}^m \otimes C_{B}^n$ ,$\{e_1,...,e_{mn}\}$ be the standard orthogonal base $\{|11>,...,|1n>,...,|m1>,...,|mn> \}$ and $\rho= \Sigma_{l=1}^{t} p_l P_{v_l}$ be a mixed state on $H$. We may consider the $mn\times t$ matrix $A$ as a $m\times 1$ blocked matrix with each block $A_w$, where $w=1,...,m$, a $n\times t$ matrix corresponding to $\{|w1>,...,|wn>\}$. Then it is easy to see $\rho_{ij}=A_iP(A_j^{*})^{\tau}$, where $i=1,...m,j=1,...,m$. Thus\\

$$
\begin{array}{cccccc}
\Sigma r_ir_j^{*} \rho_{ij}=(\Sigma r_i A_i)P(\Sigma r_i^{*} A_i^{*})^{\tau}
\end{array}
(4)
$$

{\bf Lemma 2.} {\em $\Sigma r_ir_j^{*} \rho_{ij}$ is a (semi) positive definite $n\times n$ matrix. It is singular if and only if the rank of $(\Sigma r_i A_i)$ is strictly smaller than $n$.}\\

{\bf Proof.} The first conclusion is clear. The matrix is singular if and only if there exist a vector $c=(c_1,...,c_n)$ with the property.\\

$$
\begin{array}{cccccccccc}
c(\Sigma r_ir_j^{*} \rho_{ij})(c^{*})^{\tau}=\\
(\Sigma r_i cA_i)P(\Sigma r_i^{*} c^{*}A_i^{*})^{\tau}=0
\end{array}
(5)
$$

Since $P$ is a strictly positive definite matrix,our conclusion follows immediately.\\

{\bf Proof of Theorem 2.} From Lemma 2 , we know that $V_{A}(\rho)$ is the zero locus of all $n\times n$ submatrices of $(\Sigma r_i A_i)$ in the case $t \geq n$ or the whole space $C^m$ in the case $t<n$. The conclusion is proved.\\

{\bf Remark 2.} Since the determinants of all $n \times n$ submatrices of $(\Sigma r_i A_i)$ are HOMOGENEOUS polynomials of degree $n$ , thus we can see $V_{A}(\rho)$(resp. $V_{B}(\rho)$) is an algebraic subset (called determinantal varieties in algebraic geometry [18],[19]) in $CP^{m-1}$(resp. $CP^{n-1}$).\\

Now suppose that the mixed state $\rho$ is separable,ie, there are unit product vectors $a_1 \otimes b_1,....,a_s\otimes b_s$ such that $\rho=\Sigma_{l=1}^{s}q_l P_{a_l \otimes b_l}$ , where $q_1,...q_s$ are positive real numbers. Let $a_u=a_u^1 |1>+...+a_u^m |m>,b_u= b_u^1 |1>+...+b_u^n|n>$ for $u=1,...,s$. Hence the vector representation of $a_u \otimes b_u$ with the standard base is $a_u \otimes b_u= \Sigma_{ij} a_u^ib_u^j |ij>$. Consider the corresponding $mn \times s$ matrix $C$ of $a_1 \otimes b_1,...,a_s \otimes b_s$ as in Lemma 1, we have $\rho=CQ(C^{*})^{\tau}$, where $Q$ is diagonal matrix with diagonal entries $q_1,...,q_s$. As before we consider $C$ as $m\times 1$ blocked matrix with blocks $C_w$, $w=1,...m$. Here $C_w$ is a $n \times s$ matrix of the form $C_w=(a_j^{w}b_j^i)_{1\leq i \leq n, 1 \leq j \leq s}=BT_w$ , where $B=(b_j^i)_{1\leq i \leq n, 1\leq j\leq s}$ is a $n \times s$ matrix and $T_w$ is a diagonal matrix with diagonal entries $a_1^{w},...,a_{s}^{w}$. Thus from Lemma 1, we have $\rho_{ij}=C_i Q (C_j^{*})^{\tau}=B(T_i Q (T_j^{*})^{\tau})(B^{*})^{\tau}=BT_{ij}(B^{*})^{\tau}$, where $T_{ij}$ is a diagonal matrix with diagonal entries $q_1 a_1^i(a_1^j)^{*},...,q_s a_s^i(a_s^{j})^{*}$.\\

{\bf Proof of Theorem 3.} As in the proof of Theorem 2, we have\\

$$
\begin{array}{ccccccc}
\Sigma r_ir_j^{*} \rho_{ij}=\Sigma r_i r_j^{*}BT_{ij}(B^{*})^{\tau}\\
B(\Sigma r_ir_j^{*}T_{ij})(B^{*})^{\tau}
\end{array}
(6)
$$

Here we note  $\Sigma r_ir_j^{*}T_{ij}$ is a diagonal matrix with diagonal entries\\ 
$ q_1(\Sigma r_ia_1^{i})(\Sigma r_ia_1^{i})^{*},...,q_s(\Sigma r_ia_s^{i})(\Sigma r_ia_s^{i})^{*}$.Thus $\Sigma r_ir_j^{*} \rho_{ij}=BGQ(G^{*})^{\tau}(B^{*})^{\tau}$,\\ where $G$ is a diagonal matrix with diagonal entries $ \Sigma r_ia_1^{i},...,\Sigma r_ia_s^{i}$. Because $Q$ is a strictly positive definite matrix, from lemma 2 we know that  $\Sigma r_ir_j^{*} \rho_{ij}$ is singular $n\times n$ matrix if and only if the rank of $BG$ is strictly smaller than $n$. Note that $BG$ is just the multiplication of $s$ diagonal entries of $G$ (which is linear forms of $r_1,...,r_m$) on the $s$ columns of $B$,  thus the determinants of all $n \times n$ submatrices of $BG$ (in the case $ s \geq n$, otherwise automatically linear)are the multiplications of a constant (possibly zero) and $n$ linear forms of $r_1,...,r_m$. Thus the conlusion is proved.\\

From Theorem 1, 2,3,  we can immediately have the following example of a continous family of rank 3 (entangled) mixed states ${\rho_t}$ ($t$ is a complex parameter)
on $3 \times 3$ quantum systems with the following properties:\\

1) $\rho_{t}$ is entangled (from Theorem 3) except $((t^3+2)/t)^3=0,-216,27$;\\

2) $\rho_{t}$ and $\rho_s$ is equivalent under joint unitary operation on $H=C_{A}^3 \otimes C_{B}^3$, but not equivalent if A and B just perform their local unitary operations, except $k((t^3+2)/t)=k((s^3+2)/s)$ ,where $k(x)=\frac{x^3(x^3+216)^3}{(-x^3+27)^3}$ is the moduli function.\\

{\bf Example 2.} $\rho_t=\frac{1}{3}(P_{v_1}+P_{v_2}+P_{v_3})$ where \\

$$
\begin{array}{cccccccc}
v_1=\frac{1}{\sqrt{|t|^6+2}}(t^3 |11>+|22>+|33>)\\
v_2=\frac{1}{\sqrt{3}}(|12>+|23>+|31>)\\
v_3=\frac{1}{\sqrt{3}}(|13>+|21>+|32>)
\end{array}
(7)
$$

It is easy to calculate that $\Sigma r_i A_i$ is the following $3 \times 3$ matrix\\

$$
\left(
\begin{array}{ccc}
t^3r_1&r_3&r_2\\
r_2&r_1&r_3\\
r_3&r_2&r_1
\end{array}
\right)
$$

Thus $V_{A}(\rho_t)$ is defined by $t^3r_1^3+r_2^3+r_3^3-(t^3+2)r_1r_2r_3=0$ in $CP^{2}$. With $tr_1=r_1'$ we have $r_1'^{3}+r_2^3+r_3^3-((t^3+2)/t)r_1'r_2r_3=0$. This is the well-known universal family of elliptic curves (except the special values in 1) which correspond to 3 lines, see [27]). The 2nd conclusion of 2) is from the fact about the moduli of elliptic curves ,which said that two elliptic curves are isomorphic if and only if their moduli function values are the same(see [27]). \\

We can observe that if $|t|^6=|s|^6$, the eigenvalues of $\rho_t, tr_B(\rho_t),tr_A(\rho_t)$ and $\rho_s, tr_{B}(\rho_s),tr_A(\rho_s)$ are the same, thus their von Neumann entopies are the same. However it is easy to check that the moduli fuction values at $t= 3\omega$ ($\omega$ is a primitive 3-rd root of 1) and $s=-3$ are not the same. Thus we know that $\rho_{3\omega}$ and $\rho_{-3}$ have the same spectra and local spectra (thus the same von Neumann entropies), but they are not equivalent under local unitary operations, moreover both $\rho_{3\omega}$ and $\rho_{-3}$ are entangled mixed states. This can be compared with the example in [28]. In [28] two mixed states $\rho$ and $\delta$ with the same global and local spectra are given, however $\rho$ is entangled and $\delta$ is separable.\\

Actually this example has a higher dimensional analogue of Calabi-Yau manifolds, the entangled states corresponding this family of Calabi-Yau manifolds will be studied in [24].\\

The following example is a rank 7 PPT mixed state $\rho$  on $H=C_{A}^4 \otimes C_{B}^6$. We prove it is a entangled mixed state by our Theorem 3 (thus bound entanglement). Actually our construction method is general and can be used to contruct more examples with rich algebraic and Hermitian geometric structure in their associated algebraic sets (see [24],[25]).\\

{\bf Example 3.} Consider the following 4 $ 6 \times 7$ matrices\\

$$
A_1=
\left(
\begin{array}{ccccccccccc}
1&0&0&0&0&0&0\\
0&1&0&0&0&0&0\\
0&0&1&0&0&0&0\\
0&0&0&2&0&0&1\\
0&0&0&0&2&0&0\\
0&0&0&0&0&2&0
\end{array}
\right)
$$

$$
A_2=
\left(
\begin{array}{ccccccccccc}
0&1&1&-1&0&0&1\\
1&0&1&0&0&0&0\\
1&1&0&0&0&0&0\\
-1&0&0&0&1&1&0\\
0&0&0&1&0&1&0\\
0&0&0&1&1&0&0
\end{array}
\right)
$$

$$
A_3=
\left(
\begin{array}{ccccccccccc}
2&0&0&0&0&0&0\\
0&1&1&0&0&0&0\\
0&1&1&0&0&0&0\\
0&0&0&2&0&0&0\\
0&0&0&0&1&1&0\\
0&0&0&0&1&1&0
\end{array}
\right)
$$
,and $A_4=(I_6,0)$, where $I_6$ is $6\times 6$ unitary matrix.\\

Let $A$ be a $24 \times 7$ matrix with 4 blocks $A_1,A_2,A_3,A_4$ where the 24 rows correspond to the standard base $\{|ij>\}$,$i=1,...,4,j=1,...,6$. Let $\rho$ be $(1/D)(A(A^{*})^{\tau})$ (where $D$ is a normalizing constant), a mixed state on $H$. It is easy to check that $A_i(A_j^{*})^{\tau}=A_j(A_i^{*})^{\tau}$, hence $\rho$ is invariant under partial transpose. This is a PPT mixed state.\\

As in the proof of Theorem 2 it is easy to compute $F=r_1A_1+r_2A_2+r_3A_3+r_4A_4$.\\                         

$$
\left(
\begin{array}{ccccccccccc}
u_1&r_2&r_2&-r_2&0&0&r_2\\
r_2&u_1'&r_2+r_3&0&0&0&0\\
r_2&r_2+r_3&u_1'&0&0&0&0\\
-r_2&0&0&u_2&r_2&r_2&r_1\\
0&0&0&r_2&u_2'&r_2+r_3&0\\
0&0&0&r_2&r_2+r_3&u_2'&0
\end{array}
\right)
(8)
$$
,where $u_1=r_4+r_1+2r_3,u_1'=r_4+r_3+r_1$ and $u_2=r_4+2r_3+2r_1,u_2'=r_4+2r_1+r_3$.\\

We consider the following matrix $F'$ which is obtained by adding the 7-th  column of $F$ to the 4-th column of $F$ and adding $r_2/r_1$ of the 7-th column to the 1st column.\\

$$
\left(
\begin{array}{cccccccccccc}
v_1&r_2&r_2&0&0&0&r_2\\
r_2&u_1'&r_2+r_3&0&0&0&0\\
r_2&r_2+r_3&u_1'&0&0&0&0\\
0&0&0&v_2&r_2&r_2&r_1\\
0&0&0&r_2&u_2'&r_2+r_3&0\\
0&0&0&r_2&r_2+r_3&u_2'&0
\end{array}
\right)
(9)
$$
,where $v_1=r_4+r_1+2r_3+\frac{r_2^2}{r_1},v_2=r_4+2r_3+3r_1$.\\

It is clear that the determinantal varieties defined by $F$ and $F'$ are the same in the affine chart $C^3$ with inhomogeneous coordinates $r_2'=\frac{r_2}{r_1},r_3'=\frac{r_3}{r_1},r_4'=\frac{r_4+r_3+2r_1}{r_1}$. From Theorem 3 and its proof, we know that the determinantal variety defined by $F'$ in the above affine chart (ie.$V_A(\rho) \cap C^3$)  has to be the union of some linear subset (ie., affine planes or lines in $C^3$ which may not pass the origin point) if $\rho$ is separable. We want to prove that this is impossible.\\

It is easy to check that any irreducible component(see [18]) of the algbraic set $V_{A}(\rho) \cap C^3$ cannot be dimension 2 (actually it is of dimension 1 by a proposition in p.67 of [19]). We oberseve the 1st $6 \time 6$ submatrix is a blocked diagonal matrix with 2 $3 \times 3$ blocks. Let their determinants be $f_1,f_2$ ( polynomials in the above affine  coordinates $r_2',r_3',r_4'$). It is easy to see $F'$ has rank less than 6 when $f_1=f_2=0$. Thus the algebraic set $V'=\{(r_2',r_3',r_4'): f_1=f_2=0\}$ is the sum of some irreducible components of $V_{A}(\rho) \cap C^3$. We just need to prove one of these components is not linear.\\

We have $f_2=(r_4'-r_2'-r_3')((r_3')^2+(r_4')^2-2(r_2')^2+2r_3'r_4'+r_3'r_2'+r_4'r_2'+r_2'+r_3')$. If $r_4'=r_2'+r_3'$ we have $f_1=2(r_2')^3+2(r_2')^2r_3'+4(r_3')^2-(r_2')^2+6r_2'r_3'-3r_2'-4r_3'+1$. The last polynomial $g(r_2',r_3')$ is irreducible polynomial of $r_2',r_3'$. Hence we know that one component $V_1'=\{(r_2',r_3',r_4'):r_4'=r_2'+r_3',g(r_2',r_3')=0\}$ of $V'$ is not linear. Thus $\rho$ is a entangled PPT mixed state.\\

For further results and analysis of algebraic-geometric and Hermitian-geometric properties of the algebraic sets of the mixed states in both bipartite or multipartite quantum systems and their relations with quantum entanglement, we refer to our paper [25].\\

The author acknowledges the support from NNSF China, Information Science Division, grant 69972049.\\

e-mail: dcschenh@nus.edu.sg\\

\begin{center}
REFERENCES
\end{center}

1.C.H. Bennett and P.W.Shor, Quantum Information Theory, IEEE Trans. Inform. Theory, vol.44(1998),Sep.\\

2.J.Preskill, Physics 229: Advanced mathematical methods of Physics--Quantum Computation and Information (California Institute of Technology, Pasadena,CA,1998), http://www.theory.caltech.edu/people/preskill/ph229/.\\

3.B.M.Terhal, Detecting quantum entanglements, quant-ph/0101032\\

4.R.Werner, Phys.Rev., A40,4277(1989)\\

5.A.Peres, Phys.Rev. Lett. 77,1413(1996)\\

6.M.Horodecki, P.Horodecki and R.Horodecki, Phys. Lett. A 223,8 (1996)\\

7.P.Horodecki, Phys.Lett. A 232 333(1997)\\

8.M.Horodecki , P.Horodecki and R. Horodecki, in ``Quantum Infomation---Basic concepts and experiments'', Eds, G.Alber and M.Wiener (Springer Berlin)\\

9.M. Lewenstein, D Bruss, J.I.Cirac, B.Krus, M.Kus, J. Samsonowitz , A.Sanpera and R.Tarrach, J.Mod. Optics, 47, 2481,(2000), quantu-ph/0006064\\

10.C.H.Bennett, D.P.DiVincenzo, T.Mor, P.W. Shor, J.A.Smolin and T.M. Terhal, Phys.Rev. Lett., 82: 5385-5388,1999\\

11.D.P.DiVincenzo, T.Mor, P.W.Shor, J.A.Smolin and B.M.Terhal, quant-ph/ 9908070, to appear in Comm.Math. Phys.\\

12.B.Kiraus, J.I.Cirac, S.Karnas and M.Lewenstein, quant-ph/9912010\\

13.S.Karnas and M.Lewenstein, quant-ph/0102115\\

14.P.Horodecki ,M.Lewenstein, G.Vidal and J.I. Cirac,quant-ph/0002089\\

15.J.A.Smolin, quant-ph/0001001\\

16 N.Linden and P.Popescu, Fortsch. Phys.,46, 567(1998), quant-ph/9711016\\

17.B.Terhal and P.Horodecki, Phys. Rev. A61, 040301(2000); A.Sanpera, D.Bruss and M.Lewenstein, Phys. Rev. A63 R03105(2001)\\

18.J.Harris, Algebraic geometry, A first Course, Gradute Texts in Mathematics, 133, Springer-Verlag, 1992,
  especially its Lecture 9  ``Determinantal Varieties''\\

19.E.Arbarello, M.Cornalba, P.A.Griffiths and J.Harris, Geometry of algebraic curves,Volume I, Springer-Verlag, 1985, Chapter II''Determinantal Varieties''\\ 

20.P.A.Griffiths, Infinitesimal variations of Hodges structure III:Determinantal varieties and the infinitesimal invariants of normal functions, Compositio Math. 50(1983),no.2-3, pp267-324\\

21.C.De Concini, D.Eisenbud and C.Procesi, Young diagrams and determinantal varieties, Invent. Math. ,56(1980), no.2, pp129-165\\

22.S.Abhyankar, Combinatorics of Young tableaux, determinantal varieties and the calculation of Hilbert functions(in French), Rend. Sem. Mat. Univ. Politec. Torino, 42(1984),no.3,pp65-88\\

23.Hao Chen, New invariants and separability criterion of the mixed states: multipartite case, preprint, July, 2001\\

24.Hao Chen, Families of inequivalent entangled mixed states related to Calabi-Yau manifolds, in preparation\\

25.Hao Chen, Quantum entanglement and geometry of determinantal varieties, in preparation\\

26.Hao Chen, Geometry of the determinantal varieties associated with unextendible product basis and bound entanglement, in preparation\\

27.E.Brieskorn and H.Knorrer, Plane algebraic curves, Birkhauser Boston,1981\\

28.M.A.Nielsen and J.Kemper, Phys. Rev. Lett., 86(2001), pp5184-5185

\end{document}